\documentclass[9pt,twocolumn,twoside]{osajnl}

\journal{ol} 

\setboolean{shortarticle}{true}

\makeatletter
\renewcommand{\@biblabel}[1]{#1. }
\renewcommand{\@dotsep}{500}
\renewcommand{\@pnumwidth}{0em}
\renewcommand{\l@figure}[2]{
\@dottedtocline{1}{1.5em}{2em}{Figure #1}{}\vspace{15pt}}

\newcommand{\ks}[1]{\textcolor{black}{#1}} 
\newcommand{\xl}[1]{\textcolor{black}{#1}} 
\newcommand{\xlt}[1]{\textcolor{black}{#1}} 
\usepackage[normalem]{ulem}

\begin{document}

\title{Kerr optical parametric oscillation in a photonic crystal microring for accessing the infrared}

\author[1,2,3]{Xiyuan Lu}
\author[1]{Ashish Chanana}
\author[1,2]{Feng Zhou}
\author[1]{Marcelo Davanco}
\author[1,2,*]{Kartik Srinivasan}

\affil[1]{Physical Measurement Laboratory, National Institute of Standards and Technology, Gaithersburg, MD 20899, USA}
\affil[2]{Joint Quantum Institute, NIST/University of Maryland, College Park, MD 20742, USA}
\affil[3]{e-mail: xiyuan.lu@nist.gov}
\affil[*]{Corresponding author: kartik.srinivasan@nist.gov}

\begin{abstract}
      \noindent Continuous wave optical parametric oscillation (OPO) provides a flexible approach for accessing mid-infrared wavelengths between 2~$\mu$m to 5~$\mu$m, but has not yet been integrated into silicon nanophotonics. Typically, Kerr OPO uses a single transverse mode family for pump, signal, and idler modes, and relies on a delicate balance to achieve normal (but close-to-zero) dispersion near the pump and the requisite higher-order dispersion needed for phase- and frequency-matching. Within integrated photonics platforms, this approach results in two major problems. First, the dispersion is very sensitive to geometry, so that small fabrication errors can have a large impact. Second, the device is susceptible to competing nonlinear processes near the pump. In this letter, we propose a flexible solution to infrared OPO that addresses these two problems, by using a silicon nitride photonic crystal microring (PhCR). The frequency shifts created by the PhCR bandgap enable OPO that would otherwise be forbidden. We report an intrinsic optical quality factor up to (1.2~$\pm$~0.1)$\times$10$^6$ in the 2~$\mu$m band, \ks{and use a PhCR ring to demonstrate} an OPO with threshold power of (90~$\pm$~20)~mW dropped into the cavity, with the pump wavelength at 1998~nm, and the signal and idler wavelengths at 1937~nm and 2063~nm, respectively. We further discuss how to extend OPO spectral coverage in the mid-infrared. These results establish the PhCR OPO as a promising route for integrated laser sources in the infrared.
\end{abstract}

\maketitle

\begin{figure*}[t!]
\centering\includegraphics[width=0.95\linewidth]{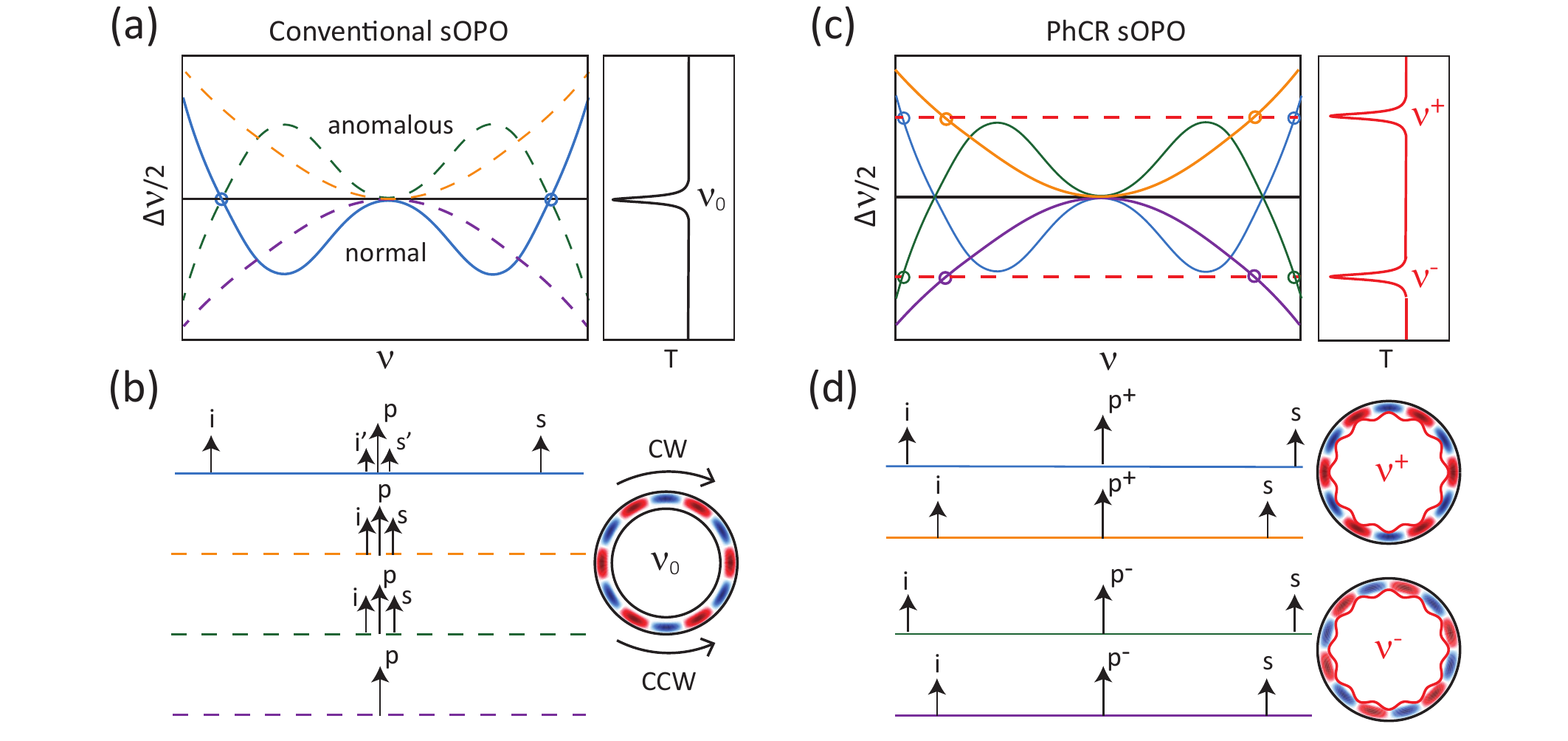}
\caption{\textbf{Motivation for Kerr OPO in a photonic crystal ring (PhCR).} \textbf{(a)} Conventional single-mode-family OPO (sOPO) requires a specific dispersion profile to create output signal and idler fields that are widely separated. The frequency mismatch is given by $\Delta \nu = \nu_s+\nu_i-2\nu_p$, where $\nu_s$, $\nu_i$, and $\nu_p$ represent signal, idler and pump frequencies, respectively. This frequency mismatch ($\Delta \nu$) is typically \ks{negative} (corresponding to a normal dispersion) but close-to-zero around the pump, \ks{and higher-order dispersion produces frequency matching ($\Delta \nu = 0$) for widely separated $\nu_s$ and $\nu_i$ (blue profile).} All other dispersion profiles (yellow, green, purple dashed lines) are not suitable for the sOPO. The right panel illustrates the transmission ($T$) of the pump mode. \textbf{(b)} The sOPO is intrinsically susceptible to close-to-pump parametric processes (blue solid line), while the other dispersion profiles (dashed lines) do not generate widely-separated OPO. Here clockwise (CW) and counter-clockwise (CCW) propagating pump modes and the \ks{resulting} OPOs behave the same. \textbf{(c)} In a PhCR resonator, the grating leads to two standing-wave modes (re-normalized from CW and CCW modes) at a targeted azimuthal mode ($m = 6$ for illustration) with a mode splitting of $\nu^{\pm} = \nu_0 \pm \beta/(2\pi)$. The frequency mismatch that would otherwise be present in a conventional microring can be compensated by this splitting, with the choice of $\nu^{+}$ or $\nu^{-}$ depending on the starting dispersion, as illustrated by two dashed red lines. \textbf{(d)} PhCR OPO works for all kinds of dispersion profiles. We show four examples with the top two using the $\nu^{+}$ mode and the bottom two using the  $\nu^{-}$ mode. Close-band OPO processes in the anomalous dispersion profiles (orange and green) are suppressed because the pump mode splitting now causes them to be frequency mismatched. \ks{We focus on the purple dispersion profile in this work.}}
\label{Fig1}
\end{figure*}

\noindent Infrared (IR) lasers at wavelengths from 2~$\mu$m to 5~$\mu$m are useful for environmental/gas sensing~\cite{Yamazoe_SABC_1994_Enviromental, Hodgkinson_MST_2012_Optical} and earth monitoring~\cite{Liu_ASR_2017_A}. \xlt{Compact sources are particularly relevant for deployable applications, and to that end, semiconductor and doped fiber lasers have been developed for wavelengths below 3~$\mu$m~\cite{Tittel_Book_2003_Mid, Jackson_NatPhoton_2012_Towards}, and interband cascade~\cite{Yang_JSTQE_2007_Distributed} and quantum cascade lasers (QCLs)~\cite{Lu_APL_2011_2.4W} have been developed for longer wavelengths. Another approach for compact mid-IR access is through optical parametric oscillation (OPO) in whispering gallery mode (WGM) resonators, with both the second-order ($\chi^{(2)}$) and third-order ($\chi^{(3)}$) nonlinearity studied~\cite{strekalov_nonlinear_2016}. In mm-size resonators, $\chi^{(2)}$ systems with wavelength access up to 8~$\mu$m~\cite{meisenheimer_continuous-wave_2017} and conversion efficiency $>10~\%$ with few mW output power has been achieved~\cite{jia_continuous-wave_2018}. Similarly, mm-scale $\chi^{(3)}$ (Kerr) resonators have realized wavelength access into the mid-infrared~\cite{Sayson_NatPhoton_2019_Octave,fujii_octave-wide_2019}. Mid-IR OPO in chip-integrated platforms is particularly compelling for scalable manufacturing and integration, and recently, Kerr OPO in AlN microrings pumped at 2~$\mu$m has been shown~\cite{Tang_OL_2020_Widely}.}

\xlt{In this letter, we demonstrate Kerr OPO in the 2~$\mu$m band, using a high quality factor ($Q$ up to 10$^6$) silicon nitride (Si$_3$N$_4$) photonic crystal ring (PhCR). The Si$_3$N$_4$ platform is of particular interest considering its widespread availability in silicon photonics foundries~\cite{lin_mid-infrared_2017}. Infrared microresonator frequency combs~\cite{luke_broadband_2015} and widely-separated OPO with pump lasers from 780~nm to 1550~nm~\cite{Lu_Optica_2019_Milliwatt, Lu_Optica_2020_On, Domeneguetti_Optica_2021_Parametric} have been shown in Si$_3$N$_4$, but thus far not into the mid-infrared with a 2~$\mu$m pump laser. One challenge at longer wavelengths is the thick Si$_3$N$_4$ film required for a suitable dispersion design. For example, AlN OPOs~\cite{Tang_OL_2020_Widely} and Si$_3$N$_4$ microcombs~\cite{luke_broadband_2015} at these wavelengths had thicknesses of $\approx$1~$\mu$m. Realizing such thicknesses in Si$_3$N$_4$ grown by the common low pressure chemical vapor deposition technique is challenging due to its large tensile stress, though mitigation techniques have enabled >700-nm-thick films~\cite{Ji_APLPhoton_2021_Methods}. In contrast, the PhCR approach we use for phase- and frequency-matching enables OPO at a 500~nm film thickness that would otherwise be unusable because of its large normal dispersion. Moreover, our approach intrinsically bypasses close-band OPO processes that are a common challenge for conventional microring Kerr OPO~\cite{Stone_PhysRevAppl_2022_Conversion}. We generate signal and idler separated by about 120~nm, and discuss how to extend this approach to wider spectral separations with idler further in the infrared. Our work establishes PhCR OPO as a promising route for chip-integrated infrared lasers.}

The typical approach for $\chi^{(3)}$ OPO in an integrated microring resonator uses a single-mode-family for all three modes in the OPO process~\cite{Lu_Optica_2019_Milliwatt, Sayson_NatPhoton_2019_Octave, Lu_Optica_2020_On,Domeneguetti_Optica_2021_Parametric}, as proposed by Lin et. al.~\cite{Lin_OE_2008_A}, and termed sOPO. This approach has the advantage of near-perfect mode overlap ($\eta$~$>$~90~\%), and using the fundamental mode family typically leads to higher optical quality factors (Qs) and smaller mode volumes (Vs) than other modes, which greatly enhances the intensity of the light fields and the power efficiency for nonlinear interaction. The sOPO approach requires a specific dispersion profile, however, to excite widely separated signal and idler. This dispersion profile should be normal but close-to-zero around the pump, as illustrated by the blue solid line in Fig.~\ref{Fig1}(a). All other dispersion profiles (dashed lines in orange, blue, and purple) in Fig.~\ref{Fig1}(a) are not suitable for generating widely-separated OPO, and their typical output spectra are illustrated in Fig.~\ref{Fig1}(b). For the specific dispersion profile that works (blue solid line), the generated sOPO is susceptible to competing close-to-pump OPO (labeled $s'$ and $i'$), though it only occurs after the excitation of the targeted widely-separated OPO (labled $s$ and $i$), due to cross-four-wave-mixing effects~\cite{Stone_PhysRevAppl_2022_Conversion}. When the dispersion is anomalous around the pump (yellow/green dashed lines), the close-to-pump OPO becomes the preferred processes. Theoretically, widely-separated OPOs are possible to generate with the green dispersion profile. In practice, however, it is difficult to achieve, as the close-to-pump modes typically have better mode overlaps and optical quality factors on average.
\begin{figure*}[t!]
\centering\includegraphics[width=0.95\linewidth]{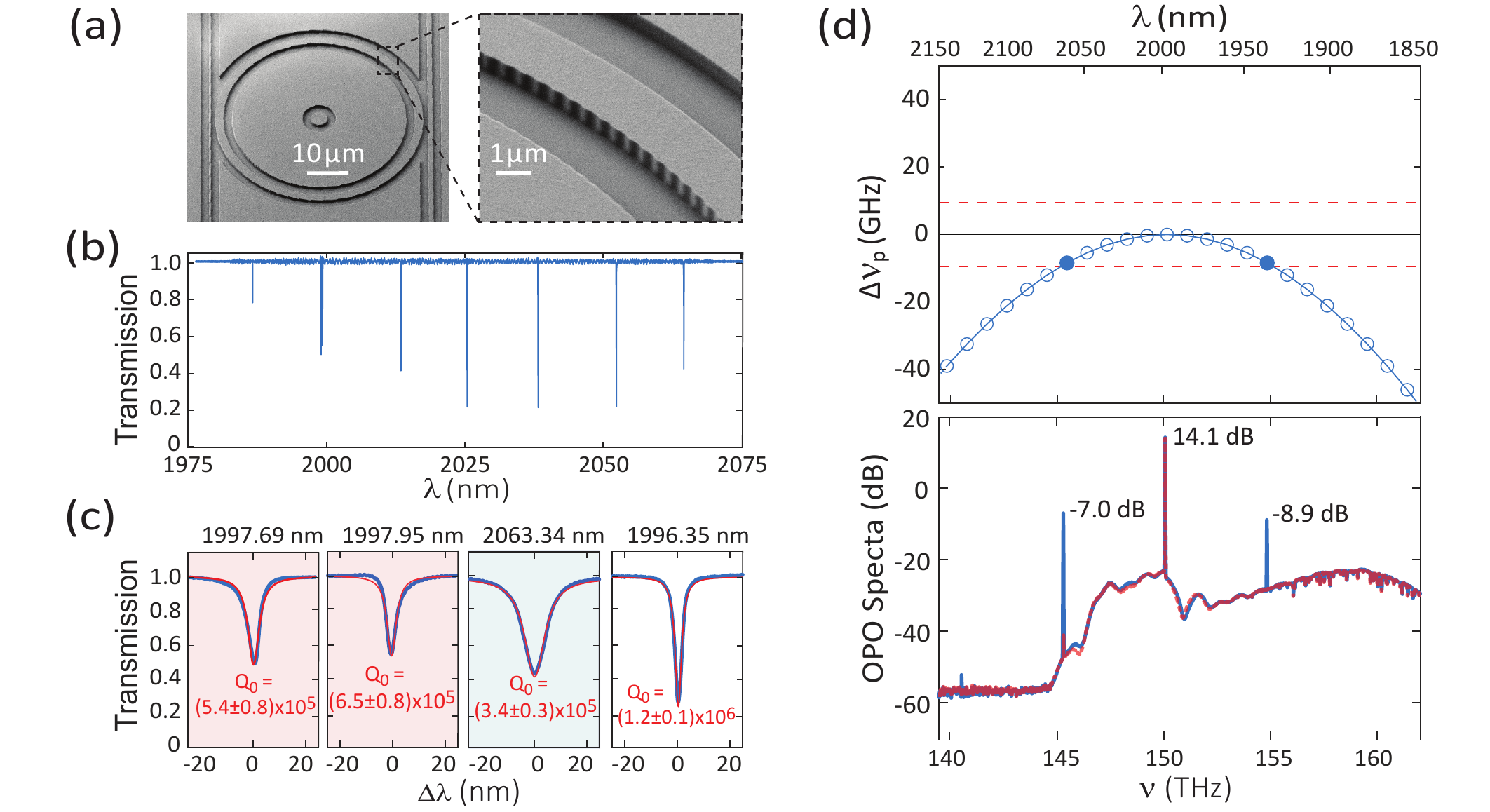}
\caption{\textbf{Experimental results of an infrared PhCR OPO.} \textbf{(a)} Scanning electron microscope images of a Si$_3$N$_4$ PhCR and a zoomed-in view of the modulation. \xlt{Only one waveguide is used in this work to couple the signal, pump, and idler modes.} \textbf{(b)} Normalized transmission for such a PhCR device showing mode splitting at approximately 1998~nm (shaded in red) and a mode without splitting at approximately 2063~nm (shaded in blue). \textbf{(c)} The three panels in red and blue show fits to the highlighted modes, and the fourth shows a high-$Q$ mode from an unmodulated device with $RW = 3~\mu$m. \textbf{(d)} The top panel shows the simulation result of the frequency matching condition with pump mode at $m = 122$ and $\nu_p$ at approximately 150~THz. The red dashed lines correspond to the pump mode with higher (the top line) and lower (the bottom line) frequencies, which correspond to the 1997.69~nm and 1997.95~nm modes in (\textbf{c}), respectively. The empty circles represent discrete mode frequencies spaced by free spectral ranges. The two solid circles are the signal and idler modes with frequency and phase matching suitable for optical parametric oscillation, while the dispersion is normal around the pump. The bottom panel shows two OPO spectra. The red dashed trace shows the spectrum near the threshold and the blue solid trace shows the spectrum above the threshold. \xlt{The weak sideband near 140~THz is likely due to a cascaded effect, as its spacing from the idler equals the idler-pump spacing.} \xl{On the y axis, 0 dB is referenced to 1 mW, i.e., dBm.}}
\label{Fig2}
\end{figure*}

In a PhCR, the grating can induce a coupling of clockwise (CW) and counter-clockwise (CCW) modes with degenerate frequencies of $\nu_0$, leading to two standing-wave modes with a mode splitting of $\nu^{\pm} = \nu_0 \pm \beta/(2\pi)$, where $\beta$ has a linear dependence on the modulation amplitude~\cite{Lu_APL_2014_Selective}. The cavity transmission and mode profiles of the targeted mode are illustrated in the right panels in Fig.~\ref{Fig1}(c)-(d), respectively. \xlt{The PhCR mode splitting only occurs in the targeted mode, which we use as the pump mode.} Remarkably, this method allows any dispersion profile for widely-separated OPO, whose signal and idler are labeled by circles in Fig.~\ref{Fig1}(c). For example, the normal dispersion profile (purple) can have widely-separated OPO with the lower-frequency pump mode, ith a frequency matching line that is effectively shifted from the black line to the bottom dashed red line. Moreover, in all these configurations, no close-to-band processes are in competition, as shown in Fig.~\ref{Fig1}(d) and in sharp contrast to Fig.~\ref{Fig1}(b), as they are now frequency mismatched due to the pump mode splitting. The scheme illustrated in purple in Fig.~\ref{Fig1}(c)-(d) has been demonstrated in the telecom previously~\cite{Black_NP_2020_Selective}, while the other three schemes have not been proposed so far. We also note that telecom ultra-low threshold OPO has been demonstrated in two-dimensional PhC defect cavities~\cite{marty_photonic_2021}, though the signal-idler separation was $<$1~THz.

To verify the proposed idea for infrared PhCR OPO, we follow the configuration illustrated in purple in Fig.~\ref{Fig1}(c)-(d) and fabricate a microring with a thickness of $H$ = 500~nm, a ring outer radius of $RR$ = 25~$\mu$m, and a ring width of $RW$ = 2~$\mu$m. The devices were fabricated in LPCVD Si$_3$N$_4$ on thermal SiO$_2$ grown on a Si wafer. The device pattern was defined by electron-beam lithography and transferred to Si$_3$N$_4$ by reactive ion etching. Figure~\ref{Fig2}(b) shows scanning electron microscope images of a fabricated PhCR with two straight waveguides, and a magnified view of a part of the microring highlighting the periodic modulation. The inner boundary of the PhCR has a sinusoidal modulation with $N$ = 122$\times$2 periods in a round trip and a modulation amplitude of $A$ = 20~nm. Such a modulation selectively splits the \xlt{fundamental transverse-electric-like} mode with $m = 122$, whose resonance is located at $\approx2~\mu$m in finite-element-method simulation. Figure~\ref{Fig2}(b) shows the normalized transmission trace in a range of $1975$~nm to $2075$~nm measured by scanning a tunable infrared laser. The transmission trace shows the PhCR resonances with a free-spectral range (FSR) of approximately 0.94~THz. The targeted mode with $m = 122$ splits into a doublet at $\lambda = 1997.69$~nm and $1997.95$~nm, with $Q_0$ of $(5.4\pm0.8)\times10^5$ and $(6.5\pm0.8)\times10^5$, respectively, as shown in Fig.~\ref{Fig2}(c), and the latter is used as the pump later on. The mode at 2063.34~nm is five FSRs away (with $m = 117$), and is used as the idler. \xlt{The discrete nature of mode resonances requires a proper mode splitting around a signal-idler mode pair, which has not been illustrated in Figure~\ref{Fig1}.} The highest $Q_0$ is $(1.2\pm0.1)\times10^6$, as shown in the last panel of Fig.~\ref{Fig2}(c), in a microring with $RW = 3~\mu$m and no modulation. The uncertainty in $Q_0$ represents the 95~\% confidence interval from nonlinear fitting.

The top panel of Fig.~\ref{Fig2}(d) shows the simulated results in open circles, for OPO frequency mismatch to the pump mode ($\Delta\nu_p$) at $\approx150$~THz, obtained from finite-element-method simulations with the same parameters as the PhCR other than the modulation. The frequency mismatch is defined as $\Delta \nu_p = (\nu_s+\nu_i)/2 - \nu_p$, where $\nu_p$, $\nu_s$ and $\nu_i$ represent the pump, signal, and idler frequencies, respectively. This definition is related to our previous definition~\cite{Lu_Optica_2019_Milliwatt, Lu_Optica_2020_On} by $\Delta \nu_p = \Delta \nu/2$, and is more convenient to use with \ks{the split pump modes of a PhCR}. $\Delta \nu_p < 0$ corresponds to a normal dispersion that typically forbids OPO generation. The top and bottom red dashed lines correspond to the 1997.69~nm and 1997.95~nm pump modes, respectively. The two solid circles represent the signal and idler modes excited by the OPO device. The bottom panel shows the OPO spectra recorded by an optical spectrum analyzer at threshold (red dashed curve) and above threshold (blue solid curve). At threshold, the power dropped into the microring is $(90\pm20)$~mW, the converted idler power is approximately $-40$ dB to the pump, and the signal is masked by the amplified spontaneous emission from the \ks{thulium}-doped fiber amplifier. Above threshold, the on-chip idler power is $(0.4\pm0.1)$~mW with the on-chip pump power of $(160\pm40)$~mW. These values are extracted from the output spectra in Fig.~\ref{Fig2}(d) and subtracting the fiber-chip insertion loss of $(3.3\pm1.3)$~dB per facet. The uncertainties are from the one-standard-deviation fluctuation in the fiber-chip coupling.

\begin{figure}[t!]
\centering\includegraphics[width=0.98\linewidth]{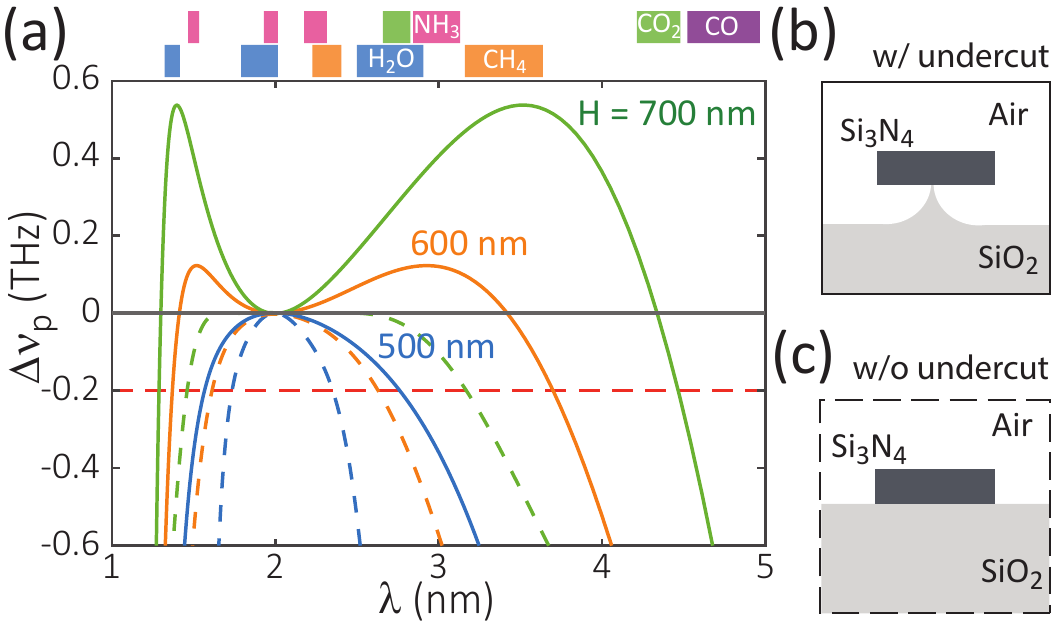}
\caption{\textbf{Design towards broader infrared spectral coverage.} \textbf{(a)} The output idler wavelength can be extended up to 4.5~$\mu$m in devices with thicker Si$_3$N$_4$ and with undercut. Here we assume a shift in pump frequency of $-$0.2~THz (the red dashed line) through the PhCR design. Beyond 3.5~$\mu$m, an undercut of the SiO$_2$ substrate is necessary to avoid absorption in SiO$_2$ and maintain high $Q$. All devices are simulated with the finite-element method using a ring width of 2~$\mu$m and thicknesses marked in the plot. The absorption ranges of several common gases are specified on the top axis for environmental gas sensing. \textbf{(b-c)} The solid and dashed curves in (a) correspond to devices with and without the undercut, whose cross-sectional structure are illustrated in (b) and (c), respectively.}
\label{Fig3}
\end{figure}

Going forward, it is of interest to extend the OPO spectrum further into the infrared. We assess the opportunities for doing so by considering widely-separated OPO designs for differing Si$_3$N$_4$ thicknesses. The dispersion curves of these structures are shown in Fig.~\ref{Fig3}(a) with thickness of 500~nm (used previously), 600~nm, and 700~nm for the blue dashed line, orange dashed line, and green dashed line, respectively. All other nominal parameters except the thickness are the same. Here we assume a -0.2~THz shift of the pump frequency, as shown by the dashed red line, which corresponds to a mode splitting of 0.4~THz, less than half of a free spectral range (0.94~THz). Assuming $\beta$ linearly depends on $A$~\cite{Lu_APL_2014_Selective}, this splitting requires $A \approx$~425~nm, approximately 21~\% of $RW$. \ks{We note that Q~$>10^6$ has been maintained with such large $A$ values in a PhCR in the telecom~\cite{Lu_NatPhoton_2022_High}. If such high-$Q$s are maintained in the infrared, we predict that these devices can generate light at around 2.4~$\mu$m, 2.6~$\mu$m, and 3.2~$\mu$m for \xl{a thickness} ($H$) of 500~nm, 600~nm, and 700~nm, respectively.}

Beyond 3.5~$\mu$m, SiO$_2$ becomes absorptive so the devices have to be undercut to minimize loss, with a device cross-section illustrated in Fig.~\ref{Fig3}(b), fabricated from the standard non-undercut cross-section in Fig.~\ref{Fig3}(c). The undercut process can be achieved using potassium hydroxide (KOH), which is known to etch only SiO$_2$ but not Si$_3$N$_4$~\cite{William_JMS_2003_Etch}.
The dispersion curves for the undercut devices are quite different and shown in the solid lines. The idler is now shifted to 2.8~$\mu$m, 3.7~$\mu$m, and 4.5~$\mu$m for $H$ of 500~nm, 600~nm, and 700~nm, respectively. 
The relevance of this potential OPO infrared spectral coverage is addressed in the top portion of Fig.~\ref{Fig3}, where we display the absorption ranges of several common gas species as examples~\cite{Rothman_JQSRT_1992_The}. 

In summary, we demonstrate infrared Kerr OPO in a PhCR. The PhCR approach enables frequency matching for a wide range of dispersion profiles, excludes competing OPO processes near the pump, and is promising for broad spectral coverage. Future tasks include pump laser integration and customized spectral emission and extension to longer infrared wavelengths.

\smallskip
\noindent \textbf{Acknowledgements.}
The authors acknowledge Paulina Kuo for instrument support and thank Carlos Ríos and Mingkang Wang for helpful discussions. This work is partly supported by the NIST-on-a-chip program and the DARPA LUMOS program.


\bibliographystyle{osajnl}
\bibliography{iOPO.bib}

\begin{thebibliography}{10}
\newcommand{\enquote}[1]{``#1''}

\bibitem{Yamazoe_SABC_1994_Enviromental}
N.~Yamazoe and N.~Miura, {\protect\JournalTitle{Sens. Actuators B: Chem.}}
  \textbf{20}, 95 (1994).

\bibitem{Hodgkinson_MST_2012_Optical}
J.~Hodgkinson and R.~P. Tatam, {\protect\JournalTitle{Meas. Sci. Technol.}}
  \textbf{24}, 012004 (2012).

\bibitem{Liu_ASR_2017_A}
C.~Liu, G.~Kirchengast, S.~Syndergaard, E.~Kursinski, Y.~Sun, W.~Bai, and
  Q.~Du, {\protect\JournalTitle{Adv. Space Res.}} \textbf{60}, 2776 (2017).

\bibitem{Tittel_Book_2003_Mid}
F.~K. Tittel, D.~Richter, and A.~Fried, \enquote{Mid-infrared laser
  applications in spectroscopy,} in \emph{Solid-state mid-infrared laser
  sources,}  (Springer, 2003), pp. 458--529.

\bibitem{Jackson_NatPhoton_2012_Towards}
S.~D. Jackson, {\protect\JournalTitle{Nat. Photon.}} \textbf{6}, 423 (2012).

\bibitem{Yang_JSTQE_2007_Distributed}
R.~Q. Yang, C.~J. Hill, K.~Mansour, Y.~Qiu, A.~Soibel, R.~E. Muller, and P.~M.
  Echternach, {\protect\JournalTitle{J. Sel. Top. Quantum Electron.}}
  \textbf{13}, 1074 (2007).

\bibitem{Lu_APL_2011_2.4W}
Q.~Y. Lu, Y.~Bai, N.~Bandyopadhyay, S.~Slivken, and M.~Razeghi,
  {\protect\JournalTitle{Appl. Phys. Lett.}} \textbf{98}, 181106 (2011).

\bibitem{strekalov_nonlinear_2016}
D.~V. Strekalov, C.~Marquardt, A.~B. Matsko, H.~G.~L. Schwefel, and G.~Leuchs,
  {\protect\JournalTitle{Journal of Optics}} \textbf{18}, 123002 (2016).

\bibitem{meisenheimer_continuous-wave_2017}
S.-K. Meisenheimer, J.~U. Fürst, K.~Buse, and I.~Breunig,
  {\protect\JournalTitle{Optica}} \textbf{4}, 189 (2017).

\bibitem{jia_continuous-wave_2018}
Y.~Jia, K.~Hanka, K.~T. Zawilski, P.~G. Schunemann, K.~Buse, and I.~Breunig,
  {\protect\JournalTitle{Optics Express}} \textbf{26}, 10833 (2018).

\bibitem{Sayson_NatPhoton_2019_Octave}
N.~L.~B. Sayson, T.~Bi, V.~Ng, H.~Pham, L.~S. Trainor, H.~G.~L. Schwefel,
  S.~Coen, M.~Erkintalo, and S.~G. Murdoch, {\protect\JournalTitle{Nat.
  Photon.}} \textbf{13}, 701 (2019).

\bibitem{fujii_octave-wide_2019}
S.~Fujii, S.~Tanaka, M.~Fuchida, H.~Amano, Y.~Hayama, R.~Suzuki, Y.~Kakinuma,
  and T.~Tanabe, {\protect\JournalTitle{Optics Letters}} \textbf{44}, 3146
  (2019).

\bibitem{Tang_OL_2020_Widely}
Y.~Tang, Z.~Gong, X.~Liu, and H.~X. Tang, {\protect\JournalTitle{Opt. Lett.}}
  \textbf{45}, 1124 (2020).

\bibitem{lin_mid-infrared_2017}
H.~Lin, Z.~Luo, T.~Gu, L.~C. Kimerling, K.~Wada, A.~Agarwal, and J.~Hu,
  {\protect\JournalTitle{Nanophotonics}} \textbf{7}, 393 (2017).

\bibitem{luke_broadband_2015}
K.~Luke, Y.~Okawachi, M.~R.~E. Lamont, A.~L. Gaeta, and M.~Lipson,
  {\protect\JournalTitle{Opt. Lett.}} \textbf{40}, 4823 (2015).

\bibitem{Lu_Optica_2019_Milliwatt}
X.~Lu, G.~Moille, A.~Singh, Q.~Li, D.~A. Westly, A.~Rao, S.-P. Yu, T.~C.
  Briles, S.~B. Papp, and K.~Srinivasan, {\protect\JournalTitle{Optica}}
  \textbf{6}, 1535 (2019).

\bibitem{Lu_Optica_2020_On}
X.~Lu, G.~Moille, A.~Rao, D.~A. Westly, and K.~Srinivasan,
  {\protect\JournalTitle{Optica}} \textbf{7}, 1417 (2020).

\bibitem{Domeneguetti_Optica_2021_Parametric}
R.~R. Domeneguetti, Y.~Zhao, X.~Ji, M.~Martinelli, M.~Lipson, A.~L. Gaeta, and
  P.~Nussenzveig, {\protect\JournalTitle{Optica}} \textbf{8}, 316 (2021).

\bibitem{Ji_APLPhoton_2021_Methods}
X.~Ji, S.~Roberts, M.~Corato-Zanarella, and M.~Lipson,
  {\protect\JournalTitle{APL Photon.}} \textbf{6}, 071101 (2021).

\bibitem{Stone_PhysRevAppl_2022_Conversion}
J.~R. Stone, G.~Moille, X.~Lu, and K.~Srinivasan, {\protect\JournalTitle{Phys.
  Rev. Appl.}} \textbf{17}, 024038 (2022).

\bibitem{Lin_OE_2008_A}
Q.~Lin, T.~J. Johnson, R.~Perahia, C.~P. Michael, and O.~J. Painter,
  {\protect\JournalTitle{Opt. Express}} \textbf{16}, 10596 (2008).

\bibitem{Lu_APL_2014_Selective}
X.~Lu, S.~Rogers, W.~C. Jiang, and Q.~Lin, {\protect\JournalTitle{Appl. Phys.
  Lett.}} \textbf{105}, 151104 (2014).

\bibitem{Black_NP_2020_Selective}
J.~A. Black, S.-P. Yu, and S.~B. Papp, {\protect\JournalTitle{Advanced
  Photonics Congress}} \textbf{OSA}, NpW2E.1 (2020).

\bibitem{marty_photonic_2021}
G.~Marty, S.~Combrié, F.~Raineri, and A.~De~Rossi, {\protect\JournalTitle{Nat.
  Photonics}} \textbf{15}, 53 (2021).

\bibitem{Lu_NatPhoton_2022_High}
X.~Lu, A.~McClung, and K.~Srinivasan, {\protect\JournalTitle{Nat. Photon.}}
  \textbf{16}, 66 (2022).

\bibitem{William_JMS_2003_Etch}
K.~Williams, K.~Gupta, and M.~Wasilik, {\protect\JournalTitle{{J
  Microelectromech. Syst.}}} \textbf{12}, 761 (2003).

\bibitem{Rothman_JQSRT_1992_The}
L.~Rothman, R.~Gamache, H.~Tipping, C.~Rinsland, M.~Smith, D.~Benner, V.~Devi,
  J.~Flaud, C.~Camy-Peyret, A.~Perrin, A.~Goldman, S.~Massie, L.~Brown, and
  R.~Toth, {\protect\JournalTitle{J. Quantum Spectrosc. Radiat. Transf.}}
  \textbf{48}, 469 (1992).

\end{thebibliography}


\end{document}